\documentclass[%
preprint,
superscriptaddress,
reprint,
 amsmath,amssymb,
 aps,
 twocolumn
]{revtex4-2}

\usepackage{graphicx}
\usepackage{dcolumn}
\usepackage{bm}
\usepackage[bookmarks=false,
 breaklinks=true,pdfborder={0 0 0},pdfborderstyle={},backref=false,colorlinks=true]
 {hyperref}
\usepackage{mathtools}
\usepackage{physics}
\usepackage{amsmath}
\usepackage{bm}
\usepackage{CJKutf8}
\usepackage{subcaption}
\usepackage[justification=raggedright, singlelinecheck=false]{caption}
\usepackage{xcolor}
\usepackage{booktabs}
\usepackage{multirow}
\usepackage{array}
\usepackage{hhline}
\usepackage{braket}
\usepackage{todonotes}

\definecolor{softred}{HTML}{B54A4A}
\definecolor{softblue}{HTML}{386FA4}

\begin{document}

\title{Wave packet landscape for open quantum systems}

\author{Kang Xu (\begin{CJK}{UTF8}{gbsn}许康\end{CJK})}
\affiliation{Graduate School of China Academy of Engineering Physics, Beijing 100193,
China}

\author{Miao-Miao Yi (\begin{CJK}{UTF8}{gbsn}易淼淼\end{CJK})}

\author{Zi-Hong Yan (\begin{CJK}{UTF8}{gbsn}燕紫鸿\end{CJK})}

\author{C. P. Sun  (\begin{CJK}{UTF8}{gbsn}孙昌璞\end{CJK})}
\email{suncp@gscaep.ac.cn}

\affiliation{Graduate School of China Academy of Engineering Physics, Beijing 100193,
China}

\date{\today}

\begin{abstract}
We formulate a landscape theory for the wave packet evolution of harmonically trapped particles with quantum fluctuations and related dissipation. We show that the long-time fates of wave packets including diffusion, localization, and collapse depend on the symmetries of an underlying landscape in covariance space. In the quantum landscape approach presented in this paper, the trapping potential and bath fluctuation break the landscape symmetry in distinct ways: the former lifts the valley-shaped landscape of a fluctuation-free free particle into a bowl-shaped landscape, leading to collapse, whereas the latter tilts the valley and turns localization into diffusion. The resulting landscape symmetry breaking accounts for the noncommuting long-time limits and abrupt changes in the asymptotic wave packet width. This establishes landscape symmetry breaking as a unified geometric origin of wave packet diffusion, localization, and collapse in quantum Brownian motion.
\end{abstract}

\maketitle

\section{Introduction}
The spreading of quantum wave packets is a fundamental manifestation of quantum mechanics, but its long-time fate in an open environment remains subtle~\cite{einstein2005born,PhysRevA.49.592,PhysRevA.51.1845,Unruh1989}. In closed systems a free wave packet spreads indefinitely, while a trapping potential can stabilize its width as a coherent state in harmonic trap. In open systems~\cite{BreuerPetruccione2002,CallenWelton1951}, quantum dissipation and fluctuations modify this simple picture: a wave packet may diffuse, remain localized, or collapse, depending on the interplay between damping, noise, and confinement~\cite{Quantum_Dissipative_Systems}. These possibilities are familiar in quantum Brownian motion~\cite{CALDEIRA1983374,CALDEIRA1983587,GRABERT1988,Hu1992}, decoherence theory~\cite{Unruh1989}, and the quantum-to-classical transition~\cite{Zurek2003}, yet they are usually discussed as separate dynamical regimes rather than as different outcomes of a single organizing principle.

The closest classical analog in classical mechanics is the potential-energy landscape~\cite{Wright1932,Frank1984,FrankScience1995,Wales1997,Wales2001,Wales2018,Onuchic1997,Wang2019}. The extrema of a potential determine equilibrium configurations and their stability. For open systems, however, the dynamics is generally not Hamiltonian, and a conventional potential-like function may not exist. Nevertheless, many dissipative systems can still be understood through an effective landscape or Lyapunov-like function~\cite{LYAPUNOV01031992,Graham1984} that organizes their long-time behavior~\cite{Ao2004,pnas_Ao,Wang2008,Wang2012}. The question addressed here is whether an analogous geometric structure exists for the spreading of quantum wave packets, where the relevant dynamical variables are not only mean positions but also variances and covariances.

In this paper we show that such a geometric structure emerges in linear open quantum dynamics. For free and harmonically trapped particles coupled to a dissipative environment, the position and momentum variances, together with their covariances, form a closed set of linear equations. From this closed dynamics, we construct a landscape in covariance space. The asymptotic width of the wave packet is then determined by the geometry of the ground manifold of this landscape, including its minima and flat directions.

This viewpoint reveals a simple geometric origin of several apparently distinct long-time behaviors. For a fluctuation-free dissipative free particle, the covariance landscape takes the form of a valley with translational symmetry. The downhill motion along the landscape stops on this degenerate ground manifold, leading to wave packet localization with a final width fixed by the initial covariance. A harmonic trap breaks this landscape symmetry by lifting the valley into a single bowl which drives the wave packet to collapse. Consequently, the long-time wave packet width changes discontinuously between the exactly free case and the limit of vanishing trap frequency. Thus long-time limit and the limit of vanishing trap do not commute.

Environmental fluctuations break the same landscape symmetry in a qualitatively different way. Instead of lifting the valley into a confining bowl, they tilt or shift the landscape. For a free particle this tilt converts localization into diffusion; for a trapped particle it shifts the minimum and produces a finite stationary width. The abrupt changes of the asymptotic wave packet width are therefore not accidental singularities of explicit solutions. They reflect how trap-induced and fluctuation-induced perturbations act as distinct sources of landscape symmetry breaking.

The landscape formulation also extends to coupled dissipative harmonic oscillators. In that case the long-time covariance is controlled by the ground manifold of the landscape and conserved quantities, without constructing an explicit scalar landscape. This establishes landscape symmetry breaking as a common geometric mechanism behind dissipation-induced localization, trap-induced collapse, and fluctuation-induced diffusion in open quantum wave packet dynamics.

\section{Landscape for wave packet spreading in dissipation}
We consider $N$ coupled harmonic oscillators linearly interacting with an Ohmic bosonic bath, namely,  the Caldeira--Leggett model~\cite{CALDEIRA1983374,CALDEIRA1983587,Ford1988}, since a broad class of open quantum systems can be reduced to this model near equilibrium. The system then obeys the quantum Langevin equation
\begin{equation}
    \ddot{\bm q}+\Gamma \dot{\bm q}+\Omega \bm q=\bm \xi(t),
    \label{linear_dissipative_q}
\end{equation}
where $\bm{q} = (q_1, \dots, q_N)^T$ is the coordinate vector, $\Gamma$ is the dissipation matrix, and $\Omega$ is the Hessian matrix of the external field at equilibrium. The fluctuation operator $\bm{\xi}(t)$ is a colored noise with zero mean $\braket{\bm{\xi}(t)}=0$. It satisfies the fluctuation-dissipation relation $\braket{\{\xi_i(t),\xi_j(t')\}}=2\hbar\Gamma_{ij}\mathcal{K}(t-t')/\pi$, where  $\mathcal{K}(\tau)=\int_0^\infty\omega\coth(\beta\hbar\omega/2)\cos(\omega\tau) d\omega$~\cite{Fundamental_aspects}. It should be noted that $\bm\xi(t)$ does not commute with the operators $\bm q(t)$ and $\bm p(t)=\dot{\bm q}(t)$, which is crucial to preserve the canonical commutation relations $[q_i,p_j]=i\hbar\delta_{ij}$.

The spread of a wave packet in configuration space can be characterized by its position variances $\{\Delta_{q_i}=\ev{q_i^2}-\ev{q_i}^2\}$, which are diagonal entries of the covariance matrix
\begin{equation}
    \Sigma\equiv\frac{1}{2}\ev{\Delta \bm{x}\Delta\bm{x}^T+(\Delta \bm{x}\Delta \bm{x}^T)^T}.
    \label{covariance_matrix}
\end{equation}
Here $\bm{x}=(\bm{q}^T,\bm{p}^T)^T$ and $\Delta \bm{x}=\bm{x}-\ev{\bm{x}}$. It follows from Eq.~\eqref{linear_dissipative_q} that the covariances obey closed equations of motion (EOM)
\begin{equation}
    \dot{\Sigma}=H\Sigma+\Sigma H^T+\Xi,
    \label{cov_lyapunov}
\end{equation}
with
\begin{equation}
H\equiv
\begin{pmatrix}
0&I\\
-\Omega&-\Gamma
\end{pmatrix},
\label{H}
\end{equation}
The inhomogeneous term $\Xi$, generated by the fluctuation, is fully determined by the symmetrized two-time correlator of the bath noise (see Appendix~\ref{A1} for specific expression). Thus, regardless of whether the noises are dominated by thermal fluctuation at high temperature or by quantum fluctuation at low temperature, the difference of their influences on wave packet spreading lies solely on their two-time correlation functions.

The landscape approach provides a straightforward geometric picture of the covariance dynamics. First we vectorize the covariance matrix by introducing $\bm{\sigma }= \operatorname{vec}(\Sigma)$, where $\operatorname{vec}(\cdot)$ denotes the standard column-stacking operation that reshapes a matrix into a single column vector by stacking its columns. Eq.~\eqref{cov_lyapunov} then becomes
\begin{equation}
    \dot{\bm{\sigma}}=H_\sigma\bm{\sigma}+\bm{\zeta}(t),
    \label{vec_eom}
\end{equation}
where $H_\sigma=I\otimes H+H\otimes I$ and $\bm{\zeta}(t)=\text{vec}(\Xi)$. Since $\Sigma$ is symmetric, the vector $\bm{\sigma}$ contains only $N(2N+1)$ independent variables. One may remove this redundancy by half-vectorization, but the full vectorization applied here keeps the notation simple in the following discussions. In the long-time limit $t\to\infty$ the fluctuation induced term $\zeta(t)$ converges to a time-independent constant. Under the physically relevant conditions that $\Gamma$ is positive definite and $\Omega$ is positive semidefinite, the dynamics of $\bm\sigma$ can be written in a generalized gradient form (the proof is given in Appendix~\ref{A2}),
\begin{equation}
    \dot{\bm{\sigma}}=-M\nabla \mathcal{L}(\bm{\sigma}),
    \label{vec_landscape}
\end{equation}
where $M$ is a positive-definite matrix, namely, $\bm x^T M \bm x>0$ for any nonzero vector $\bm x$. The landscape $\mathcal{L}$ is a scalar function of the covariance variable $\bm{\sigma}$~\cite{Ao2004,pnas_Ao}, which takes the quadratic form $\mathcal{L}=\bm{\sigma}^TL\bm{\sigma}/2+\bm{F}^T\bm{\sigma}$ with $L=-M^{-1}H_\sigma$ and $\bm{F}=-M^{-1}\bm{\zeta}$. In this representation, one has $\dot{\mathcal{L}}\le0$, where the equality holds only at stationary points satisfying $\nabla\mathcal{L}=0$, owing to the positive definiteness of $M$. Therefore, the dynamics in covariance space is driven toward lower values of the landscape, and the minima of $\mathcal{L}$ determine the asymptotic structure of the wave packet.

\section{Landscape for a dissipative harmonic oscillator and wave packet localization}\label{Sec_III}
To illustrate how wave-packet spreading maps onto the motion of a covariance state on the landscape, we first consider a regime in which the initial wave-packet width is much larger than the fluctuation-induced broadening. The early and intermediate evolution is then dominated by deterministic dissipative dynamics, while the fluctuation term can be neglected to leading order~\cite{PhysRevA.49.592,PhysRevA.51.1845}. The system is thus approximated by a damped harmonic oscillator, whose EOM in position $q$ is written as
\begin{equation}
    \ddot{q} + \gamma \dot{q} + \omega^2 q = 0.
    \label{osc_second_order}
\end{equation}
As discussed above, wave packet spreading is characterized by the covariances of $q$ and $p=\dot{q}$, $\Delta_q(t) = \ev{q^2} - \ev{q}^2$, $\Delta_p(t) = \ev{p^2} - \ev{p}^2$, and $\Delta_{qp}(t) = \ev{\{q,p\}}/2 - \ev{q} \ev{p}$. Note that the operator $p$ is not canonical momentum conjugate to $q$, since $\frac{d}{dt}[q,p] = -\gamma[q,p]$ due to neglect of the fluctuation. Therefore, the usual uncertainty relation between position and momentum does not apply here directly. The quantization of dissipative systems is subtle and has long been discussed in terms of effective canonical variables and effective Hamiltonians~\cite{kanai1948quantization,caldirola1941forze,peng1980quantum,Fujikawa1992,PhysRevA.49.592,PhysRevA.51.1845}. Instead of relying on an effective Hamiltonian for the dissipative motion, our landscape approach characterizes the evolution directly in covariance space, which extends straightforwardly to coupled harmonic oscillators and to quantum Brownian motion with fluctuations, as discussed below.

It follows from Eq.~\eqref{vec_eom} that the covariances obey
\begin{equation}
    \dot{\bm{\sigma}}=H_{1D}\bm{\sigma},\qquad H_{1D}=
    \begin{pmatrix}
        0 & 0 & 2\\
        0 & -2\gamma & -2\omega^2\\
        -\omega^2 & 1 & -\gamma
    \end{pmatrix},
    \label{sigma_equation}
\end{equation}
where $\bm{\sigma}(t)=(\Delta_q,\Delta_p,\Delta_{qp})^T$. This covariance dynamics is not generated by a Hamiltonian since the three-dimensional covariance space does not admit a canonical symplectic structure. Therefore the landscape introduced below cannot be viewed as a Hamiltonian, but as a Lyapunov-like topographical function in covariance space. 

We now construct the landscapes separately for a free particle ($\omega=0$) and a trapped particle ($\omega\neq0$). The landscape for Eq.~\eqref{sigma_equation} is not unique. Although a unified expression can be constructed for arbitrary $\omega$, separate constructions yield much simpler expressions and, more importantly, make their distinct geometric structures explicit.

\begin{figure}[t]
    \centering
    \includegraphics[width=8.5cm]{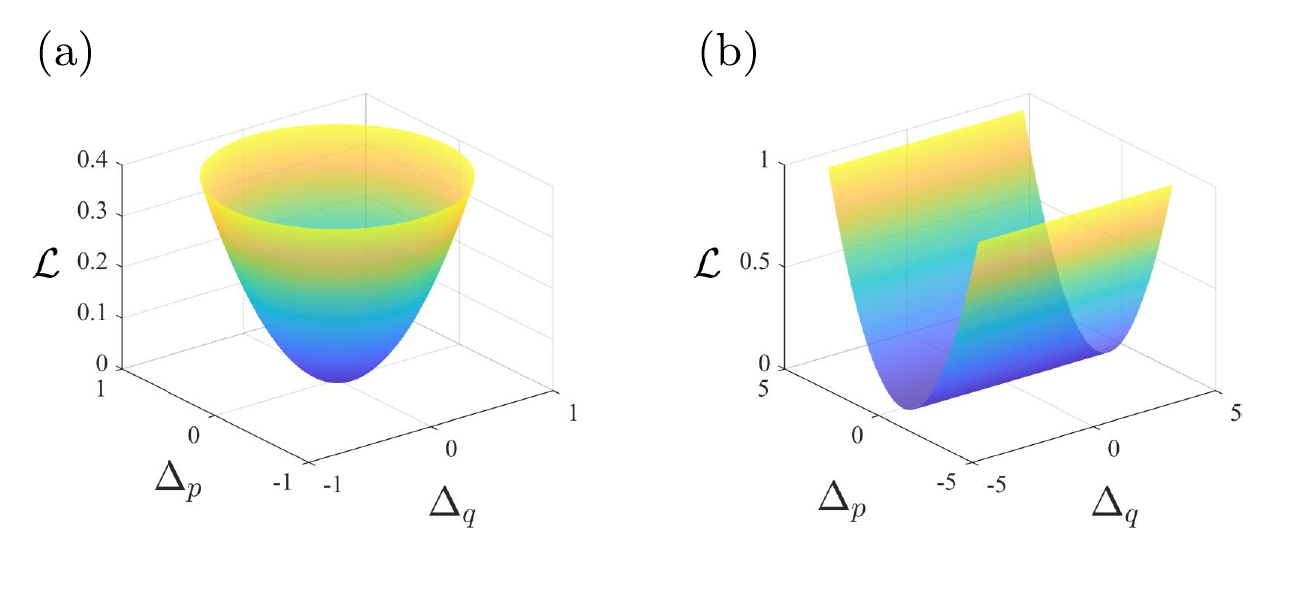}
    \caption{
\textbf{Landscape deformation by trap potential in fluctuation-free dynamics.}
The landscapes are shown in the $(\Delta_q,\Delta_p)$ projection, with the remaining covariance fixed at its minimizing value. The axes are extended to negative values of $\Delta_q$ and $\Delta_p$ only for visualizing the landscape geometry, although physical variances are restricted to be nonnegative.
(a) For $\omega\neq0$, the fluctuation-free landscape is bowl-shaped and has a unique minimum, implying collapse of the wave packet in the long-time limit. The parameters are set as $\omega=\gamma=1$.
(b) For $\omega=0$, the fluctuation-free landscape develops a flat valley of degenerate minima, reflecting translational symmetry in covariance space and allowing localization to a finite width determined by the initial condition. The parameter is set as $\gamma=1$.
}
    \label{fig_osc_landscape}
\end{figure}

For $\omega \neq 0$, one convenient decomposition is
\begin{equation}
    M =\frac{4}{\gamma(5\gamma^2+48\omega^2)}
    \begin{pmatrix}
        1 & 0 & -\frac{2\omega^2}{\gamma}  \\
        0 & 5\omega^4 & \frac{2\omega^4}{\gamma} \\
        \frac{2\omega^2}{\gamma}  & -\frac{2\omega^4}{\gamma}  & \frac 12\omega^2
    \end{pmatrix},
    \label{osc_SA_nonzeroomega}
\end{equation}
together with the quadratic landscape
\begin{align}
    \mathcal{L}_{\text{CHO}}=&\,
    \frac{5}{2}\gamma^2\omega^2
    \left[
    \Delta_q
    +\frac{\gamma}{2\omega^2}\Delta_{qp}
    -\frac{1}{5\omega^2}\Delta_p
    \right]^2 \notag\\
    &+\frac{\gamma^2(5\gamma^2+48\omega^2)}{8\omega^2}\,\Delta_{qp}^2
    +\frac{\gamma^2(5\gamma^2+48\omega^2)}{20\omega^4}
    \Delta_p^2.
    \label{L_nonflu_neq0}
\end{align}
This is a bowl-shaped landscape whose unique stable minimum is $\sigma'_m = (0,0,0)^T$, as shown in Fig.~\ref{fig_osc_landscape}. Physically, this means that any non-zero trapping potential drives the wave packet into a $\delta$ function~\cite{YU1995167}, which is referred to as wave packet collapse in the following:
\begin{equation}
    \lim_{t\to\infty}\Delta_q(t) = 0,
    \label{asymptotic_width_collapse}
\end{equation}
independent of the initial condition.

For $\omega = 0$, the trap is removed, the system reduces to a dissipative free particle~\cite{PhysRevA.49.592,PhysRevA.51.1845}. In this case, we find
\begin{equation}
    M =
    \begin{pmatrix}
        \frac{2}{\gamma^3} & 0 & -\frac{2}{\gamma^2} \\
        0 & 2\gamma & 0 \\
        0 & -1 & \frac 1\gamma
    \end{pmatrix},
    \label{osc_SA_zeroomega}
\end{equation}
with the corresponding landscape
\begin{equation}
    \mathcal{L}_{\text{FP}}=\frac{1}{2}\Delta_p^2+\frac{1}{2}\gamma^2\Delta_{qp}^2.
    \label{L_nonflu_0}
\end{equation}
Unlike the case $\omega\neq0$, $\mathcal{L}$ is now invariant under translations along the $\Delta_q$ direction. Its minima therefore form the bottom of a continuous valley rather than an isolated point, as shown in Fig.~\ref{fig_osc_landscape}.
This translational symmetry has a direct dynamical consequence. The wave packet no longer collapses with vanishing width, but may relax to any point on the bottom of the valley, determined by initial conditions. For $\omega=0$, the matrix $H_{1D}$ has a left zero mode $\bm n^T=(1,1/\gamma^2,2/\gamma)$ satisfying $\bm n^T H_{1D}=0$. Hence the quantity $\bm n^T\bm\sigma(t)$ is conserved during the time evolution (see Lemma 1 in Appendix~\ref{A3}). Since the valley of the landscape in Eq.~\eqref{L_nonflu_0} is characterized by $\Delta_p=\Delta_{qp}=0$, this conservation law gives $\lim_{t\to\infty}\Delta_q(t)=\bm{n}^T\bm{\sigma}(0)$. 

As an example, for an initial Gaussian wave packet $\psi(q)=(2\pi a^2)^{-1/4}\exp[-q^2/(4a^2)]$, one has
\begin{equation}
    \lim_{t\to\infty}\Delta_q(t)
    =
    a^2+\frac{\hbar^2}{4\gamma^2a^2}.
    \label{asymptotic_width}
\end{equation}
Thus, for damped free particle with bath fluctuation neglected, the asymptotic width of the wave packet approaches a finite non-zero value, compared with the evolution of free particle where the wave packet spreads indefinitely. This result of wave packet localization is consistent with Ref.~\cite{PhysRevA.51.1845}, where the dissipative free particle is described by the Caldirola--Kanai Hamiltonian~\cite{caldirola1941forze,kanai1948quantization}.

It is discovered by Eqs.~\eqref{asymptotic_width_collapse} and \eqref{asymptotic_width} that the asymptotic width is discontinuous between the confined limit and the exactly free case:
\begin{equation}
    \lim_{\omega\to0}\lim_{t\to\infty}\Delta_q(t)
    \neq
    \lim_{t\to\infty}\lim_{\omega\to0}\Delta_q(t).
    \label{noncommuting_limit_confinement}
\end{equation}
This non-commutativity implies a symmetry breaking of the landscape. For $\omega=0$, the landscape shapes as a translationally symmetric valley, while the long-time state is determined by the start point on the landscape, namely, the initial conditions. For any infinitesimal trapping potential, however, this degeneracy is lifted. Consequently, the translational symmetry is broken, with the shape of the landscape turning from valley to bowl, indicating the collapse of the wave packet as $t\to\infty$.

The one-dimensional example reveals that the flat directions of the landscape control the asymptotic wave packet width. This relation is not special to a single oscillator, but extends directly to coupled dissipative harmonic oscillators:

\textit{Proposition---}
Under the physical conditions where $\Omega$ is symmetric positive semidefinite and $\Gamma$ is symmetric positive definite, for the fluctuation-free dynamics governed by Eq.~\eqref{linear_dissipative_q} with $\bm{\xi}=0$, or equivalently Eq.~\eqref{vec_eom} with $\bm{\zeta}=0$, the long-time covariance vector is fixed by
\begin{equation}
    \lim_{t\to\infty}\bm{\sigma}(t)
    =
    M_r(M_lM_r)^{-1}M_l\bm{\sigma}(0).
    \label{asymptotic_covariance_general}
\end{equation}
Here $M_r=(\bm{r}_1,\cdots,\bm{r}_{d^2})$ and $M_l=(\bm{l}_1,\cdots,\bm{l}_{d^2})^T$ are constructed from the right and left zero modes of $H_\sigma=I\otimes H+H\otimes I$, denoted by $\{\bm{r}_k|\,H_\sigma\bm{r}_k=0\}$ and $\{\bm{l}_k|\,\bm{l}^T_kH_\sigma=0\}$, respectively. These zero modes can be constructed from the zero modes  $\{\bm{y}_i\}_{i=1}^{d}$ of the trap matrix $\Omega$ directly, where $\Omega\bm{y}_i=0$. Explicitly, it can be shown that $\bm{r}_k=\bm{h}_i\otimes\bm{h}_j$ and $\bm{l}_k=\tilde{\bm{h}}_i\otimes\tilde{\bm{h}}_j$ where
\begin{equation}
    \bm{h}_i=
    \begin{pmatrix}
        \bm{y}_i\\
        \bm{0}
    \end{pmatrix},
    \qquad
    \tilde{\bm{h}}_i=
    \begin{pmatrix}
        \Gamma\bm{y}_i\\
        \bm{y}_i
    \end{pmatrix},
\end{equation}
and we have relabeled the pair $(i,j)$ by a single index $k=1,\dots,d^2$.

The derivation of this proposition requires the following lemmas (see Appendix~\ref{A3} for proofs).

\textit{Lemma 1.---}
The quantities $M_l\bm{\sigma}$ are conserved under the fluctuation-free covariance dynamics. Namely,
\begin{equation}
    M_l\bm{\sigma}(t)=M_l\bm{\sigma}(0).
    \label{left_zero_mode_conservation}
\end{equation}
Equivalently, each left zero mode $\bm{l}_k$ defines a conserved quantity $\bm{l}_k^T\bm{\sigma}(t)$.

\textit{Lemma 2.---}
In the long-time limit, the covariance vector lies on the flat valley of the landscape and can be expanded as
\begin{equation}
    \lim_{t\to\infty}\bm{\sigma}(t)=\sum_k^{d^2}c_k\bm{r}_k=M_r\bm{c},
    \label{right_zero_mode_expansion}
\end{equation}
where $\bm{c}=(c_1,\dots,c_{d^2})^T$. Each vector $\bm{r}_k$ gives a translationally invariant direction of the landscape, namely, for $\forall \bm{c}\in\mathbb{R}^{d^2}$ one has $\mathcal{L}(\bm{\sigma}+M_r\bm{c})=\mathcal{L}(\bm{\sigma})$. Therefore, the hyperplane spanned by these vectors is the valley floor of the covariance landscape. 

Since we use full vectorization, the above zero modes are written in the enlarged vector space. On the physical subspace of symmetric covariance matrices, the independent flat directions are obtained by taking the symmetrized combinations of $h_i\otimes h_j$, equivalently by using half-vectorization.

This result shows that the long-time wave packet width can be obtained from the symmetry of the landscape solely. The right zero modes determine the flat valley, while the left zero modes provide the conserved quantities that select the final point on this valley. Thus, even without explicitly constructing the landscape function, its symmetry fixes the long-time state.

\section{Landscape for quantum Brownian motion}\label{Sec_QBM}
We now restore the fluctuation term and examine how it deforms the landscape obtained in the fluctuation-free case. For one-dimensional quantum Brownian motion (QBM), the particle obeys the quantum Langevin equation
\begin{equation}
    \ddot q+\gamma \dot q+\omega^2 q=\xi(t).
    \label{qbm_second_order}
\end{equation}
Correspondingly, the EOM of covariance vector $\bm\sigma=(\Delta_q,\Delta_p,\Delta_{qp})^T$ is written as
\begin{equation}
    \dot{\bm\sigma}=H_{1D}\bm\sigma+\bm\zeta .
    \label{qbm_cov_vector}
\end{equation}
Here $\bm\zeta=(0,2\Delta_{p\xi},\Delta_{q\xi})^T$, $\Delta_{q\xi}=\ev{\{q,\xi\}}/2-\ev{q}\ev{\xi}$ and $\Delta_{p\xi}=\ev{\{p,\xi\}}/2-\ev{p}\ev{\xi}$. These two quantities are completely determined by the symmetrized two-time correlator of the bath fluctuation $\ev{\{\xi(t'),\xi(t)\}}$ (see Appendix~\ref{A4} for specific expression).

An important cancellation occurs at this point. In an Ohmic bath without an ultraviolet cutoff, quantities $\Delta_{q\xi}$ and $\Delta_{p\xi}$ are not separately finite in low temperature. The position spreading, however, is not governed by them independently, but by the combination
\begin{equation}
    \gamma^2 D_\omega(T)
    \equiv
    \lim_{t\to\infty}
    \left[
        \gamma\Delta_{q\xi}(t)+\Delta_{p\xi}(t)
    \right],
    \label{finite_fluctuation_combination}
\end{equation}
which remains finite both for $\omega\neq0$ and the free-particle case $\omega=0$ as shown below. Thus the ultraviolet-sensitive parts of $\Delta_{q\xi}$ and $\Delta_{p\xi}$ cancel in the quantity that characterizes wave packet spreading in the configuration space. However, we must note that the spreading in the momentum space is characterized by $\Delta_{p\xi}$ alone, which remains cutoff dependent. The subscript in $D_\omega(T)$ indicates that the finite value depends on the trapping strength, where for the free particle we write $D_0(T)$.

For a trapped particle with $\omega\neq0$, bath fluctuations preserve the bowl shape of the fluctuation-free landscape in Eq.~\eqref{L_nonflu_neq0}, but shift its minimum. Introducing the shifted covariance variables
\[
\widetilde{\Delta}_q
=
\Delta_q-\frac{\gamma\Delta_{q\xi}+\Delta_{p\xi}}{\gamma\omega^2},
\qquad
\widetilde{\Delta}_p
=
\Delta_p-\frac{\Delta_{p\xi}}{\gamma},
\]
the landscape becomes
\begin{equation}
    \mathcal{L}_{\text{QBM}}
    =
    \mathcal{L}_{\text{CHO}}
    (\widetilde{\Delta}_q,\widetilde{\Delta}_p,\Delta_{qp}).
\end{equation}
The wave packet of the dissipative oscillator therefore no longer collapses to a $\delta$-function-shaped packet. Instead, it approaches a finite width given by the landscape minimum $\tilde{\Delta}_q=\tilde{\Delta}_p=\Delta_{qp}=0$:
\begin{equation}
    \lim_{t\to\infty}\Delta_q(t)
    =
    \frac{\gamma}{\omega^2}D_\omega(T).
    \label{qbm_confined_width}
\end{equation}

In the high-temperature limit, the bath fluctuation reduces to classical white noise, $\braket{\{\xi(t),\xi(t')\}}=4\gamma k_B T\delta(t-t')$, which yields $\Delta_{q\xi}=0$ and $\Delta_{p\xi}=\gamma k_BT$ (see Appendix~\ref{A4} for the calculation). Thus we have $D_\omega(T)=k_BT/\gamma$, and
\begin{equation}
    \lim_{t\to\infty}\Delta_q(t)
    =
    \frac{k_BT}{\omega^2},
    \label{qbm_highT_confined}
\end{equation}
The asymptotic width in configuration space is therefore independent of the dissipation strength $\gamma$ in the high-temperature limit. 

At low temperature, the bath noise is governed by quantum fluctuations. This colored noise gives
\begin{equation}
     D_\omega(T)
    =
    D_\omega(0)
    +
    \frac{\pi k_B^2 T^2}{3\hbar\omega^2}+
        \frac{2\pi^3(2\omega^2-\gamma^2)k_B^4T^4}{15\hbar^3\omega^6}
        +\cdots,
        \label{D_omega_lowT}
\end{equation}
where $D_\omega(0)$ is the zero-temperature contribution from the quantum background of the bath, the rest terms give the higher order corrections at low temperature. Explicitly, one finds (see Appendix~\ref{A4} for derivation)
\begin{widetext}
\begin{equation}
D_\omega(0)
=
\frac{\hbar\omega^2}{\pi}
\begin{cases}
\displaystyle
\frac{2}{\gamma\sqrt{4\omega^2-\gamma^2}}
\arctan
\left(
\frac{\sqrt{4\omega^2-\gamma^2}}
{\gamma}
\right),
&
\gamma<2\omega,
\\[2.0em]
\displaystyle
\frac{1}{2\omega^2},
&
\gamma=2\omega,
\\[1.4em]
\displaystyle
\frac{1}{\gamma\sqrt{\gamma^2-4\omega^2}}
\ln
\left(
\frac{
\gamma+\sqrt{\gamma^2-4\omega^2}
}{
\gamma-\sqrt{\gamma^2-4\omega^2}
}
\right),
&
\gamma>2\omega ,
\end{cases}
\label{end_Domega_zeroT}
\end{equation}
\end{widetext}
where the three branches correspond to the underdamped, critically damped, and overdamped regimes, respectively. At zero temperature, the asymptotic wave packet becomes monotonically narrower as either the dissipation strength $\gamma$ or the trap frequency $\omega$ is increased, with the other parameter held fixed, throughout the entire parameter range, as follows from Eqs.~\eqref{qbm_confined_width} and \eqref{end_Domega_zeroT}.

\begin{figure}[t]
    \centering
    \includegraphics[width=8.5cm]{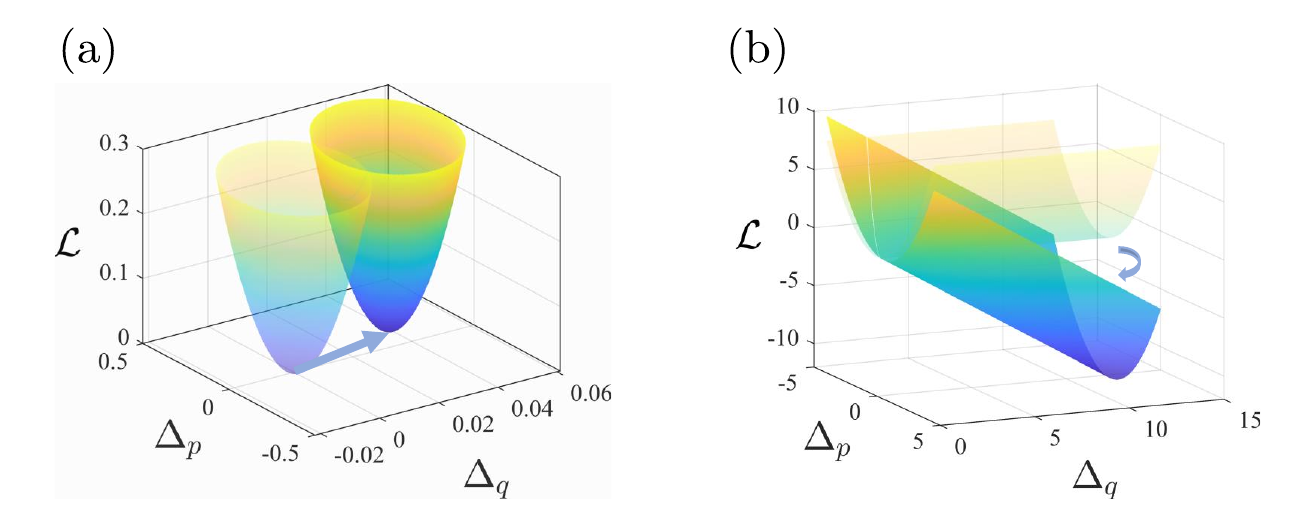}
    \caption{
\textbf{Landscape deformation by fluctuation.}
The landscapes are shown in the $(\Delta_q,\Delta_p)$ projection,
with the remaining covariance fixed at its minimizing value.
The translucent surfaces denote the fluctuation-free landscapes,
whereas the solid surfaces denote the landscapes deformed by
bath fluctuations. The parameters are set as $\gamma=1$, $\Delta_{q\xi}=\Delta_{p\xi}=0.5$. (a) In the confined case, the fluctuation
preserves the bowl topology but shifts its minimum away from
the origin, leading to a stationary wave packet with finite width.
(b) In the free-particle case, the fluctuation adds a linear term
along the flat direction, tilts the fluctuation-free valley, removes
the degenerate minimum manifold, and drives diffusive spreading,
$\Delta_q(t)\sim 2D(T)t$. 
}
    \label{fig_osc_landscape_flu}
\end{figure}

The case of quantum brown free motion (QBFM) with $\omega=0$ is qualitatively different. Bath fluctuations tilts the horizontal valley-shaped landscape given by Eq.~\eqref{L_nonflu_0}, as shown in Fig.~\ref{fig_osc_landscape_flu}. The resulting landscape is written as
\begin{equation}
\begin{aligned}
    \mathcal L_{\text{QBFM}}
    =&
    \frac12\gamma^2\left[\Delta_{qp}-D_0(T)\right]^2
    +
    \frac12\left(\Delta_p-\frac{\Delta_{p\xi}}{\gamma}\right)^2\\
    &-
    \gamma^3D_0(T)\Delta_q ,
\end{aligned}
    \label{qbm_landscape_0}
\end{equation}
where $D_0(T)$ is given by Eq.~\eqref{finite_fluctuation_combination} with $\omega=0$. The last term breaks the translational degeneracy of the fluctuation-free landscape Eq.~\eqref{L_nonflu_0} by tilting the valley along the $\Delta_q$ direction. As a result, the ground manifold of the landscape is reduced from a one-dimensional line to a zero-dimensional point at infinity in covariance space. Correspondingly, the wave packet no longer relaxes to a point with finite covariances at the bottom of the valley , but instead drifts toward arbitrarily large $\Delta_q$. In the long-time limit,
\begin{equation}
    \lim_{t\to\infty}\dot{\Delta}_q(t)=2D_0(T).
    \label{qbm_diffusion}
\end{equation}
Remarkably, one finds $D_0(T)= k_BT/\gamma$ for any nonzero temperature (see Appendix~\ref{A4} for derivation), which combining with Eq.~\eqref{qbm_diffusion} recovers the classical Einstein law
\begin{equation}
\label{linear_diffusion}
    \Delta_q(t)=\frac{2k_BT}{\gamma}t,\qquad t\to\infty,
\end{equation}
even at a low temperature. However, at zero temperature, the linear diffusion coefficient vanishes. In fact, one has $\left(\gamma\Delta_{q\xi}(t)+\Delta_{p\xi}(t)\right)/\gamma^2\approx k_BT/\gamma+\hbar/(\pi\gamma t)$ in the long-time limit, where the temperature dependent term is the leading term for any non-zero temperature. At zero temperature, the leading term vanished, the first-order correction must then be considered, which gives $\dot{\Delta}_q(t)\approx2\hbar/(\pi\gamma t)$ as $t\to\infty$. Thus the leading long-time spreading is governed by the tail of the bath correlator and becomes logarithmic~\cite{Quantum_Dissipative_Systems,GRABERT1988},
\begin{equation}
    \Delta_q(t)\sim \frac{2\hbar}{\pi\gamma}\ln(\gamma t)+\text{const}.
    \label{ln_spreading}
\end{equation}
This result also has a simple physical interpretation with the landscape approach. At zero temperature, before taking the limit \(t\to\infty\), the leading nonzero term is \(D_0(0)\simeq\hbar/(\pi\gamma t)\) (see Eq.~\eqref{zero_expansion}). Landscape Eq.~\eqref{qbm_landscape_0} then becomes a time-dependent landscape in which the tilted valley gradually approaches a horizontal one,
\begin{equation}
\begin{aligned}
    \mathcal L'_{\text{QBFM}}
    =&
    \frac12\gamma^2\left[\Delta_{qp}-D_0(T)\right]^2
    +
    \frac12\left(\Delta_p-\frac{\Delta_{p\xi}}{\gamma}\right)^2\\
    &-
    \frac{\hbar\gamma^2}{\pi t}\Delta_q .
\end{aligned}
\end{equation}
Geometrically, this landscape interpolates between a valley with a fixed slope, which yields a linear diffusion given by Eq.~\eqref{linear_diffusion}, and a horizontal valley, which corresponds to wave-packet localization in the absence of bath fluctuations. The spreading of the wave packet should therefore also interpolates between linear diffusion and localization. At zero temperature, the slope of the landscape along $\Delta_q$-axis decays as \(1/t\), giving rise to a persistent but progressively weaker driving along the valley. The accumulation of this driving over time leads to the logarithmic spreading of the wave packet.

This geometric picture also clarifies why thermal and quantum fluctuations lead to different asymptotic laws without changing the basic bowl-versus-valley distinction. In the confined case, the finite fluctuation combination in Eq.~\eqref{finite_fluctuation_combination} shifts the minimum and produces a finite stationary width. In the free-particle case, the same combination tilts the flat direction and drives unbounded spreading. The cancellation between the ultraviolet-sensitive parts of $\Delta_{q\xi}$ and $\Delta_{p\xi}$ is not a technical accident, but the reason why the landscape deformation relevant to position spreading remains well defined.

The singular role of fluctuation can be expressed through a noncommuting limit. We introduce a formal parameter $\lambda$ by replacing $\bm{\zeta}$ with $\lambda\bm{\zeta}$. For a physical bath, the fluctuation and dissipation are constrained by the fluctuation-dissipation relation; here $\lambda$ is introduced only to probe the singular perturbation of the landscape. For $\omega=0$ and $\lambda=0$, the landscape has a flat ground-state manifold, and the wave packet can retain a finite width fixed by the initial condition. For any nonzero $\lambda$, however, this manifold is tilted and pushed to infinity. Consequently,
\begin{equation}
    \lim_{\lambda\to0}\lim_{t\to\infty}\Delta_q(t)
    \neq
    \lim_{t\to\infty}\lim_{\lambda\to0}\Delta_q(t).
    \label{qbm_limit_noncommutativity}
\end{equation}
This noncommutativity is a manifestation of landscape symmetry breaking. An arbitrarily weak fluctuation breaks the translational symmetry of the fluctuation-free ground-state manifold and converts a horizontal valley into a tilted one.

\section{Conclusion}
We have formulated a landscape approach for the covariance dynamics of open quantum systems. In this representation, the long-time behavior of a wave packet is controlled by the geometry of an landscape in covariance space.  For dissipative oscillators without fluctuations, trapping potential lifts the translationally invariant flat valley of the free-particle limit into a single bowl, producing a singular difference between $\omega=0$ and $\omega\to0$. For coupled systems, the asymptotic covariance is fixed by the zero modes of the landscape and the associated conserved quantities, without requiring an explicit construction of the landscape. When bath fluctuations are included, the confined oscillator retains a shifted minimum and reaches a finite stationary width, whereas the free particle acquires a tilted valley and diffuses. Thus, the noncommuting limits associated with weak trap and weak fluctuation both originate from the zero-mode structure of the covariance landscape. This provides a compact geometric description of localization, diffusion, and fluctuation-induced degeneracy lifting in linear quantum Brownian dynamics.

\section*{Acknowledgments}
This work was supported by the National Natural Science Foundation of China (NSFC), No. 12688201 and Science Challenge Project, No.TZ2025017.

\appendix

\section{General expression for the fluctuation-induced term}
\label{A1}
In this section, we derive the general form of the inhomogeneous term
$\Xi(t)$ induced by the bath fluctuation in Eq.~\eqref{cov_lyapunov}, or equivalently
$\zeta(t)=\operatorname{vec}\Xi(t)$ in Eq.~\eqref{vec_eom}, based on the quantum Langevin equation~\eqref{linear_dissipative_q}. By introducing
\begin{equation}
    x=(\bm q^T,\bm p^T)^T,\qquad
    \eta(t)=(0,\bm\xi^T(t))^T,
\end{equation}
Eq.~\eqref{linear_dissipative_q} is written as
\begin{equation}
    \dot x=Hx+\eta(t),\qquad
    H=
    \begin{pmatrix}
        0&I\\
        -\Omega&-\Gamma
    \end{pmatrix}.
\end{equation}
Denote the retarded Green matrix as $G(t)$, which satisfies
\begin{equation}
    \ddot G(t)+\Gamma \dot G(t)+\Omega G(t)=I\delta(t),
    \qquad G(t<0)=0 .
\end{equation}
The formal solution of $\Delta x\equiv x-\braket{x}$ is then written as
\begin{equation}
    \Delta x(t)=e^{Ht}\Delta x(0)
    +\int_0^t ds\,K(t-s)\Delta\eta(s),
    \label{formal_solution}
\end{equation}
where $\Delta\eta\equiv\eta-\braket{\eta}$, and the kernel matrix $K(\tau)$ is given by
\begin{equation}
    K(\tau)=\Theta(\tau)
    \begin{pmatrix}
        0&G(\tau)\\
        0&\dot G(\tau)
    \end{pmatrix},
\end{equation}
with $\Theta(\tau)$ being the Heaviside step function.

For a factorized initial state of the system and bath, $\rho(0)=\rho_{\rm S}(0)\otimes\rho_{\rm B}(0)$, one finds that the homogeneous part $\exp(Ht)\Delta x(0)$ in Eq.~\eqref{formal_solution} does not contribute to the final expression of $\Xi(t)$. Defining the symmetrized correlation matrix of the noise,
\begin{equation}
    C_\eta(t,s)
    \equiv
    \frac{1}{2}\left\langle
    \{\Delta\eta(t),\Delta\eta(s)\}
    \right\rangle ,
\end{equation}
where the anticommutator of two vector operators $\bm{u},\bm{v}$ is defined as $\{\bm{u},\bm{v}\}_{ij}\equiv \bm{u}_i \bm{v}_j+\bm{v}_j \bm{u}_i $. One then has
\begin{equation}
    \Xi(t)
    =
    \int_0^t ds\,
    \left[
        C_\eta(t,s)K^T(t-s)
        +
        K(t-s)C_\eta(s,t)
    \right].
    \label{Xi_general}
\end{equation}
Since $C_\eta(s,t)=C_\eta^T(t,s)$, the matrix $\Xi(t)$ is symmetric, as required by the covariance dynamics.

In the Caldeira--Leggett model with an Ohmic bath, the correlation matrix $C_\eta(t,s)$ takes the following expression,
\begin{equation}
    C_\eta(t,s)
    =
    \frac{\hbar}{\pi}
    \begin{pmatrix}
        0&0\\
        0&\Gamma
    \end{pmatrix}
    \int_0^\infty d\nu\,
    \nu\coth\left(\frac{\beta\hbar\nu}{2}\right)
    \cos[\nu(t-s)] .
    \label{Ceta_CL}
\end{equation}
Using formula $\operatorname{vec}(ABC)=(C^T\otimes A)\operatorname{vec}B$, Eq.~\eqref{Xi_general} gives the vectorized inhomogeneous term
\begin{equation}
\begin{aligned}
    \zeta(t)
    =&
    \int_0^t ds\,
    \left[
        K(t-s)\otimes I_{2N}
    \right]\operatorname{vec} C_\eta(t,s)\\
    &+
    \int_0^t ds\,
    \left[
        I_{2N}\otimes K(t-s)
    \right]\operatorname{vec} C_\eta(s,t).
\end{aligned}
    \label{zeta_general}
\end{equation}
Note that $C_\eta(t,s)=C_\eta(t-s)=C_\eta(s,t)$ according to Eq.~\eqref{Ceta_CL}, one has
\begin{equation}
    \zeta(t)
    =
    \int_0^t d\tau\,
    \left[
        K(\tau)\otimes I_{2N}
        +
        I_{2N}\otimes K(\tau)
    \right]\operatorname{vec} C_\eta(\tau).
    \label{zeta_stationary}
\end{equation}
Equations~\eqref{Xi_general}--\eqref{zeta_stationary} show that the bath affects the covariance dynamics only through the symmetrized two-time correlation function of the bath noise.

For $N=1$, Eq.~\eqref{Xi_general} reduces to
\begin{equation}
    \Xi(t)=
    \begin{pmatrix}
        0&\Delta_{q\xi}(t)\\
        \Delta_{q\xi}(t)&2\Delta_{p\xi}(t)
    \end{pmatrix},
\end{equation}
where
\begin{equation}
    \Delta_{q\xi}(t)
    =
    \frac{1}{2}\int_0^t ds\,
    G(t-s)\left\langle\{\xi(s),\xi(t)\}\right\rangle ,
\end{equation}
and
\begin{equation}
    \Delta_{p\xi}(t)
    =
    \frac{1}{2}\int_0^t ds\,
    \dot G(t-s)\left\langle\{\xi(s),\xi(t)\}\right\rangle .
\end{equation}
These are the fluctuation-induced terms in Sec.~\ref{Sec_QBM} discussing the landscape for one-dimensional quantum Brownian motion.

\section{Existence of the landscape decomposition for $H_\sigma$}
\label{A2}
In this section we prove that, for $H_\sigma=I\otimes H+H\otimes I$ where
\begin{equation}
H=
\begin{pmatrix}
0 & I\\
-\Omega & -\Gamma
\end{pmatrix},
\end{equation}
providing that $\Omega$ is a symmetric positive semidefinite matrix and $\Gamma$ is a symmetric positive definite matrix, there exists a scalar function $\mathcal L(\bm{\sigma})$ such that
\begin{equation}
\dot{\bm{\sigma}}=H_\sigma \bm{\sigma}=-M\nabla \mathcal L(\bm{\sigma}),
\end{equation}
where $M$ is positive definite matrix in the sense that
\begin{equation}
\bm{v}^T M \bm{v}>0,\qquad \forall\,\bm{v}\neq 0.
\end{equation}
Here, $M$ is not required to be symmetric.

\paragraph*{Spectral property of $H$.---} 
Let $(\bm{q},\bm{p})^T$ be an eigenvector of $H$ with eigenvalue $\lambda$. Then
\begin{equation}
\bm{p}=\lambda \bm{q},
\qquad
(\lambda^2 I+\lambda \Gamma+\Omega)\bm{q}=0.
\end{equation}
Multiplying the second equation by $\bm{q}^\dagger$ from the left gives
\begin{equation}
\lambda^2 \|\bm{q}\|^2+\lambda\,\bm{q}^\dagger\Gamma \bm{q}
+\bm{q}^\dagger \Omega \bm{q}=0.
\end{equation}
Since $\Gamma> 0$ and $\Omega\ge 0$ (we denote "positive definite" and "positive semidefinite" by "$>$" and "$\ge$" respectively in the following), one finds that the real part of every nonzero eigenvalue of $H$ is negative. The eigenvectors with zero eigenvalue of $H$ satisfy
\begin{equation}
\Omega \bm{q}=0,\qquad \bm{p}=0.
\end{equation}
Thus, the dimension of the null space of $H$ is $d=\dim \ker \Omega$, where $\text{ker}\Omega$ denotes the null space (or equivalently, kernel space) of matrix $\Omega$.

We next show that the zero eigenvalue of $H$ is semisimple. Suppose, to the contrary, that a nonzero zero mode $(\bm{q},0)^T$, with $\Omega\bm{q}=0$, belongs to a nontrivial Jordan chain. Then there exists a eigenvector $(\bm{u},\bm{v})^T$ satisfying
\begin{equation}
H
\begin{pmatrix}
\bm{u}\\
\bm{v}
\end{pmatrix}
=
\begin{pmatrix}
\bm{q}\\
0
\end{pmatrix}.
\end{equation}
It follows that $\bm{v}=\bm{q}$ and
\begin{equation}
-\Omega\bm{u}-\Gamma\bm{q}=0.
\end{equation}
Left-multiplying this equation by $\bm{q}^T$, and using $\Omega\bm{q}=0$ together with the symmetry of $\Omega$, gives
\begin{equation}
\bm{q}^T\Gamma\bm{q}=0.
\end{equation}
Because $\Gamma>0$, this equality implies $\bm{q}=0$, in contradiction with the assumption that $(\bm{q},0)^T$ is a nonzero zero mode. Hence, no nontrivial Jordan chain is associated with $\lambda=0$, and the zero eigenvalue is semisimple.

It follows that there exists a real invertible matrix $S$ such that
\begin{equation}
S^{-1}HS=
\begin{pmatrix}
0_d & 0\\
0 & J_H
\end{pmatrix},
\end{equation}
where all eigenvalues of $J_H$ have negative real parts.

\paragraph*{Block form of $H_\sigma$.---}
Let $T=S\otimes S$. Then
\begin{align}
T^{-1}H_\sigma T
&=
I\otimes(S^{-1}HS)+(S^{-1}HS)\otimes I
\notag\\
&=
\begin{pmatrix}
0_{d^2} & 0\\
0 & J
\end{pmatrix},
\label{eq_blockHs}
\end{align}
where all eigenvalues of $J$ have negative real parts. Thus, in the transformed coordinates
\begin{equation}
\tilde{\bm{\sigma}}=T^{-1}\bm{\sigma},
\end{equation}
the zero-mode subspace and the asymptotically decaying subspace are explicitly separated.

\paragraph*{Construction on the stable subspace.---}
Since $J$ is Hurwitz, the Lyapunov equation
\begin{equation}
J^T P+P J=-I
\label{eq_LyapunovP}
\end{equation}
admits a unique symmetric positive-definite solution $P=P^T\succ0$.

Define
\begin{equation}
\tilde K=
\begin{pmatrix}
0_{d^2} & 0\\
0 & P
\end{pmatrix},
\qquad
\tilde{\mathcal L}(\tilde{\bm{\sigma}})
=
\frac12\tilde{\bm{\sigma}}^T\tilde K\,\tilde{\bm{\sigma}},
\end{equation}
so that
\begin{equation}
\nabla_{\tilde{\bm{\sigma}}}\tilde{\mathcal L}
=
\tilde K\,\tilde{\bm{\sigma}}.
\end{equation}
We further introduce
\begin{equation}
\tilde M=
\begin{pmatrix}
I_{d^2} & 0\\
0 & -JP^{-1}
\end{pmatrix}.
\end{equation}
Equation~\eqref{eq_blockHs} then gives
\begin{equation}
\begin{pmatrix}
0_{d^2} & 0\\
0 & J
\end{pmatrix}
\tilde{\bm{\sigma}}
=
-\tilde M\,\tilde K\,\tilde{\bm{\sigma}}
=
-\tilde M\,\nabla_{\tilde{\bm{\sigma}}}\tilde{\mathcal L}.
\end{equation}
Therefore, the desired landscape decomposition is realized in the transformed basis.

\paragraph*{Positivity of $\tilde M$.---}
Although $\tilde M$ is not necessarily symmetric, it remains to show that it is positive definite in the sense of quadratic forms. The identity block already gives a positive contribution on the zero-mode subspace, so it suffices to consider the lower block. Multiplying Eq.~\eqref{eq_LyapunovP} by $P^{-1}$ from the left and the right yields
\begin{equation}
P^{-1}J^T+JP^{-1}=-P^{-2}.
\end{equation}
Accordingly, the symmetric part of $-JP^{-1}$ is
\begin{equation}
\frac{-JP^{-1}-P^{-1}J^T}{2}
=
\frac{P^{-2}}{2}\succ0.
\end{equation}
For every nonzero real vector $\bm{x}$, one therefore has
\begin{equation}
\bm{x}^T(-JP^{-1})\bm{x}
=
\bm{x}^T
\frac{-JP^{-1}-P^{-1}J^T}{2}
\bm{x}
=
\frac12\bm{x}^TP^{-2}\bm{x}
>0.
\end{equation}
Hence, $\tilde M$ is positive definite in the required sense.

\paragraph*{Transformation back to the original variables.---}
Define
\begin{equation}
K=T^{-T}\tilde K\,T^{-1},
\qquad
M=T\,\tilde M\,T^T,
\end{equation}
together with
\begin{equation}
\mathcal L(\bm{\sigma})
=
\frac12\bm{\sigma}^T K\,\bm{\sigma}.
\end{equation}
Then
\begin{equation}
\nabla_{\bm{\sigma}}\mathcal L
=
K\bm{\sigma},
\end{equation}
and
\begin{align}
H_\sigma\bm{\sigma}
&=
T
\begin{pmatrix}
0_{d^2} & 0\\
0 & J
\end{pmatrix}
T^{-1}\bm{\sigma}
\notag\\
&=
-T\tilde M\tilde K T^{-1}\bm{\sigma}
\notag\\
&=
-MK\bm{\sigma}
=
-M\nabla_{\bm{\sigma}}\mathcal L.
\end{align}
Moreover, for any $\bm{v}\neq0$,
\begin{equation}
\bm{v}^TM\bm{v}
=
(T^T\bm{v})^T\tilde M(T^T\bm{v})
>0,
\end{equation}
because $T^T\bm{v}\neq0$. Thus, $M$ is positive definite.

We have therefore established the existence of the decomposition
\begin{equation}
H_\sigma\bm{\sigma}
=
-M\nabla\mathcal L(\bm{\sigma}),
\end{equation}
where $M$ is positive definite and $\mathcal L$ is a scalar quadratic function.

\paragraph*{Remark.---}
When $\Omega\succ0$, the kernel of $H$ is absent, and hence no zero-mode block occurs in the above construction. In this case, $\mathcal L$ is strictly positive definite. When $\Omega$ has zero modes, $\mathcal L$ is positive semidefinite, and its null directions correspond to the flat directions of the landscape.

\section{Proof of the zero-mode lemmas}
\label{A3}
\textit{Lemma 1.---}
The quantities $M_l\bm\sigma(t)$ are conserved under the fluctuation-free covariance dynamics.

\textit{Proof.---}
Since each row of $M_l$ is a left zero mode of $H_\sigma$, one has
\begin{equation}
    M_lH_\sigma=0 .
\end{equation}
Thus
\begin{equation}
    \frac{d}{dt}\bigl[M_l\bm\sigma(t)\bigr]
    =
    M_l\dot{\bm\sigma}(t)
    =
    M_lH_\sigma\bm\sigma(t)
    =
    0 .
\end{equation}
Therefore
\begin{equation}
    M_l\bm\sigma(t)=M_l\bm\sigma(0).
    \label{end_left_conservation}
\end{equation}
Equivalently, each $\bm l_k^T\bm\sigma(t)$ is a conserved quantity.
\hfill$\square$

\textit{Lemma 2.---}
The long-time covariance vector lies on the flat valley of the landscape and can be written as
\begin{equation}
    \lim_{t\to\infty}\bm\sigma(t)=M_r\bm c ,
    \label{end_right_expansion}
\end{equation}
where $\bm c=(c_1,\dots,c_{d^2})^T$. The columns of $M_r$ span the translationally invariant directions of the landscape.

\textit{Proof.---}
In the fluctuation-free case, the landscape is quadratic,
\begin{equation}
    \mathcal L(\bm\sigma)
    =
    \frac12\bm\sigma^T L\bm\sigma ,
\end{equation}
with
\begin{equation}
     L=-M^{-1}H_\sigma,
    \qquad
    M\succ0,
    \qquad
     L^T= L .
\end{equation}
If $\bm r_k$ is a right zero mode of $H_\sigma$, then
\begin{equation}
    H_\sigma\bm r_k=0
    \quad\Rightarrow\quad
     L\bm r_k=0 .
\end{equation}
Hence, for any real parameter $\alpha$,
\begin{equation}
    \mathcal L(\bm\sigma+\alpha\bm r_k)=
    \frac12
    (\bm\sigma+\alpha\bm r_k)^T
     L
    (\bm\sigma+\alpha\bm r_k) \notag=\mathcal L(\bm\sigma).
\end{equation}
Thus each $\bm r_k$ is a translationally invariant direction of the landscape, and the subspace spanned by the columns of $M_r$ is the flat valley of the landscape.

In the long time limit the system must approach its steady states with $\dot{\bm{\sigma}}=0$, which gives
\begin{equation}
    H_\sigma\bm\sigma(\infty)=0 .
\end{equation}
Therefore the asymptotic states must lie in the subspace spanned by $\{\bm{r}_k\}$,
\begin{equation}
    \bm\sigma(\infty)=M_r\bm c .
\end{equation}
This proves Lemma 2.
\hfill$\square$

We now combine the two lemmas to obtain the asymptotic covariance. Taking the long-time limit of Eq.~\eqref{end_left_conservation} and using Eq.~\eqref{end_right_expansion}, we have
\begin{equation}
    M_lM_r\bm c=M_l\bm\sigma(0).
\end{equation}
It remains to show that $M_lM_r$ is invertible.

The matrix elements of $M_lM_r$ are
\begin{align}
    (M_lM_r)_{(a,b),(i,j)}
    &=
    \bm l_{ab}^T\bm r_{ij} \notag\\
    &=
    (\tilde{\bm h}_a\otimes\tilde{\bm h}_b)^T
    (\bm h_i\otimes\bm h_j) \notag\\
    &=
    (\tilde{\bm h}_a^T\bm h_i)
    (\tilde{\bm h}_b^T\bm h_j).
\end{align}
Using the definitions of $\bm h_i$ and $\tilde{\bm h}_i$, we find
\begin{equation}
    \tilde{\bm h}_a^T\bm h_i
    =
    \bm y_a^T\Gamma\bm y_i .
\end{equation}
Define the $d\times d$ matrix $G_{ai}=\bm y_a^T\Gamma\bm y_i$. Then, up to the ordering convention used to relabel the pair indices,
\begin{equation}
    M_lM_r=G\otimes G .
\end{equation}
Since $\Gamma$ is positive definite and the vectors
$\{\bm y_i\}$ are linearly independent, $G$ is positive definite. Indeed, for any nonzero vector $\bm u$,
\begin{equation}
    \bm u^TG\bm u
    =
    \left(\sum_i u_i\bm y_i\right)^T
    \Gamma
    \left(\sum_j u_j\bm y_j\right)
    >0 .
\end{equation}
Therefore $G$ is invertible, and so is $G\otimes G$. Hence $M_lM_r$ is invertible.

Solving for $\bm c$ gives
\begin{equation}
    \bm c
    =
    (M_lM_r)^{-1}M_l\bm\sigma(0).
\end{equation}
Then we obtain
\begin{equation}
    \lim_{t\to\infty}\bm\sigma(t)
    =
    M_r(M_lM_r)^{-1}M_l\bm\sigma(0).
\end{equation}
This completes the proof of the asymptotic formula.

\section{fluctuation-induced terms and diffusion coefficient}
\label{A4}
We derive the fluctuation-induced terms entering the covariance dynamics of the one-dimensional quantum Brownian motion whose EOM is written as
\begin{equation}
    \ddot q+\gamma\dot q+\omega^2 q=\xi(t).
\end{equation}
Let $G(t)$ be the retarded Green function satisfying
\begin{equation}
    \ddot G(t)+\gamma\dot G(t)+\omega^2G(t)=\delta(t),
    \qquad
    G(t<0)=0 .
\end{equation}
The operator solutions can be written as
\begin{equation}
\begin{aligned}
    q(t)&=q_h(t)+\int_0^t ds\,G(t-s)\xi(s),\\
    p(t)&=p_h(t)+\int_0^t ds\,\dot G(t-s)\xi(s),
\end{aligned}
\end{equation}
where $q_h(t)$ and $p_h(t)$ are the homogeneous solutions. For an initially factorized system-bath state, one finds
\begin{equation}
    \Delta_{q\xi}(t)
    =
    \frac12\int_0^t ds\,G(t-s)
    \ev{\{\xi(s),\xi(t)\}},
    \label{end_delta_qxi_time}
\end{equation}
\begin{equation}
    \Delta_{p\xi}(t)
    =
    \frac12\int_0^t ds\,\dot G(t-s)
    \ev{\{\xi(s),\xi(t)\}} ,
    \label{end_delta_pxi_time}
\end{equation}
where the homogeneous parts of $q(t)$ and $p(t)$ do not contribute. For an Ohmic bath with cutoff frequency $\Lambda$, the symmetrized noise correlator is
\begin{equation}
    \frac12\ev{\{\xi(s),\xi(t)\}}
    =
    \frac{\hbar\gamma}{\pi}
    \int_0^\Lambda d\nu\,
    \nu\coth\left(\frac{\beta\hbar\nu}{2}\right)
    \cos[\nu(t-s)] ,
    \label{end_ohmic_noise}
\end{equation}
where $\beta=(k_BT)^{-1}$. This shows explicitly that the fluctuation affects the covariance dynamics only through the two-time correlator of the bath noise.

For $\omega\neq0$, the long-time limit of Eqs.~\eqref{end_delta_qxi_time} and \eqref{end_delta_pxi_time} is 
\begin{equation}
   \lim_{t\to\infty} \Delta_{q\xi}(t)
    =
    \frac{\hbar\gamma}{\pi}
    \int_0^\Lambda d\nu\,
    \frac{
    \nu(\omega^2-\nu^2)
    \coth\left(\frac{\beta\hbar\nu}{2}\right)}
    {(\omega^2-\nu^2)^2+\gamma^2\nu^2},
    \label{end_delta_qxi_freq}
\end{equation}
and
\begin{equation}
    \lim_{t\to\infty}\Delta_{p\xi}(t)
    =
    \frac{\hbar\gamma^2}{\pi}
    \int_0^\Lambda d\nu\,
    \frac{
    \nu^3
    \coth\left(\frac{\beta\hbar\nu}{2}\right)}
    {(\omega^2-\nu^2)^2+\gamma^2\nu^2}.
    \label{end_delta_pxi_freq}
\end{equation}
If there is no cutoff frequency, $\Delta_{q\xi}$ and $\Delta_{p\xi}$ are separately ultraviolet divergent. However, it is discovered from Eq.~\eqref{finite_fluctuation_combination} that
\begin{equation}
 D_\omega(T)
    =
    \frac{\hbar\omega^2}{\pi}
    \int_0^\infty d\nu\,
    \frac{
    \nu
    \coth\left(\frac{\beta\hbar\nu}{2}\right)}
    {(\omega^2-\nu^2)^2+\gamma^2\nu^2}
    \label{end_Domega_integral}
\end{equation}
is finite.  In the low-temperature limit, $k_BT\ll\hbar\omega$, by decomposition
\begin{equation}
    \coth\left(\frac{\beta\hbar\nu}{2}\right)
    =
    1+
    \frac{2}{e^{\beta\hbar\nu}-1}.
\end{equation}
 one has
\begin{equation}
    D_\omega(T)
    =
    D_\omega(0)
    +
    \frac{\pi}{3\hbar\omega^2\beta^2}+
        \frac{2\pi^3(2\omega^2-\gamma^2)}{15\hbar^3\omega^6\beta^4}
        +\cdots.
    \label{end_Domega_lowT}
\end{equation}
The first term gives the zero-temperature quantum contribution, and the rest terms give the thermal correction.

The zero-temperature contribution depends on the damping regime:
\begin{widetext}
\begin{equation}
D_\omega(0)
=
\frac{\hbar\omega^2}{\pi}
\begin{cases}
\displaystyle
\frac{2}{\gamma\sqrt{4\omega^2-\gamma^2}}
\arctan
\left(
\frac{\sqrt{4\omega^2-\gamma^2}}
{\gamma}
\right),
&
\gamma<2\omega,
\\[2.0em]
\displaystyle
\frac{1}{2\omega^2},
&
\gamma=2\omega,
\\[1.4em]
\displaystyle
\frac{1}{\gamma\sqrt{\gamma^2-4\omega^2}}
\ln
\left(
\frac{
\gamma+\sqrt{\gamma^2-4\omega^2}
}{
\gamma-\sqrt{\gamma^2-4\omega^2}
}
\right),
&
\gamma>2\omega .
\end{cases}
\label{end_Domega_zeroT}
\end{equation}
\end{widetext}
This shows that the divergent parts cancel in the physically relevant quantity $D_\omega(T)$.

In the free-particle case $\omega=0$, one has
\begin{equation}
\label{zero_expansion}
\begin{aligned}
        \frac{\gamma\Delta_{q\xi}(t)+\Delta_{p\xi}(t)}{\gamma^2}=&\frac{\hbar}{\pi\gamma}\int_0^t d\tau\int_0^\infty d\nu \nu\coth(\frac{\beta\hbar\nu}{2})\cos(\nu\tau)\\
        \approx&\frac{k_BT}{\gamma}+\frac{\hbar}{\pi\gamma}\frac 1t.
\end{aligned}
\end{equation}
which gives the logarithm spreading in Eq.~\eqref{ln_spreading} at zero temperature.

\section{An example of coupled oscillators with unequal damping}
\label{app:coupled_example}

Consider two damped particles connected by a spring, with different damping strengths and no external trap:
\begin{equation}
\begin{aligned}
\ddot q_1+\gamma_1\dot q_1+\kappa(q_1-q_2)&=0,\\
\ddot q_2+\gamma_2\dot q_2+\kappa(q_2-q_1)&=0,
\end{aligned}
\label{eq:coupled_example_eom}
\end{equation}
where $\kappa>0$ and $\gamma_1,\gamma_2>0$. In the notation of Eq.~\eqref{linear_dissipative_q}, we have
\begin{equation}
\bm q=
\begin{pmatrix}q_1\\q_2\end{pmatrix},
\qquad
\Omega=\kappa
\begin{pmatrix}
1&-1\\
-1&1
\end{pmatrix},
\qquad
\Gamma=
\begin{pmatrix}
\gamma_1&0\\
0&\gamma_2
\end{pmatrix}.
\label{eq:coupled_example_matrices}
\end{equation}
For $\gamma_1\ne\gamma_2$, the matrices $\Omega$ and $\Gamma$ do not commute, which means this system can not be decoupled into two independent systems through unitary transformation.

Let $\bm p=\dot{\bm q}$ and $\bm x=(q_1,q_2,p_1,p_2)^T$. According to Eq.~\eqref{covariance_matrix}, the covariance matrix is written as
\begin{equation*}
    \Sigma=\begin{pmatrix}
\Delta_{q_1},
\Delta_{q_1q_2},
\Delta_{q_1p_1},
\Delta_{q_1p_2}\\
\Delta_{q_1q_2},
\Delta_{q_2},
\Delta_{q_2p_1},
\Delta_{q_2p_2}\\
\Delta_{q_1p_1},
\Delta_{q_2p_1},
\Delta_{p_1},
\Delta_{p_1p_2}\\
\Delta_{q_1p_2},
\Delta_{q_2p_2},
\Delta_{p_1p_2},
\Delta_{p_2}
\end{pmatrix}.
\end{equation*}
 Column-stacking the covariance matrix gives
\begin{equation}
\bm{\sigma}=\operatorname{vec}(\Sigma).
\label{eq:coupled_example_sigma}
\end{equation}
The off-diagonal covariances appear twice because the main text uses full vectorization of the symmetric
matrix $\Sigma$.

The matrix $\Omega$ has one zero mode, $\bm y=(1,1)^T$. According to the construction below
Eq.~\eqref{asymptotic_covariance_general}, the
corresponding right and left zero modes of $H$ are
\begin{equation}
\bm h=
\begin{pmatrix}1\\1\\0\\0\end{pmatrix},
\qquad
\widetilde{\bm h}=
\begin{pmatrix}\gamma_1\\\gamma_2\\1\\1\end{pmatrix}.
\label{eq:coupled_example_h}
\end{equation}
Lemma 2 in Sec.~\ref{Sec_III} identifies the translational invariant direction of the covariance landscape as
\begin{equation}
\bm r=\bm h\otimes\bm h
=\operatorname{vec}
\begin{pmatrix}
1&1&0&0\\
1&1&0&0\\
0&0&0&0\\
0&0&0&0
\end{pmatrix}.
\label{eq:coupled_example_r}
\end{equation}
Then we have
\begin{equation}
    \mathcal L(\bm{\sigma}+a\bm r)
=\mathcal L(\bm{\sigma})
\end{equation}
for any real number $a$. Specifically, moving
along this direction changes $\Delta_{q_1}$, $\Delta_{q_2}$, and $\Delta_{q_1q_2}$ by the same amount, while leaving all position--momentum and momentum covariances unchanged. The landscape is then a hyper-valley with 1-dimensional ground manifold of degenerate minima. Therefore, one can tell directly that the wave packets in configuration space of the two particles are both localized with finite width in the far future.

Lemma 1 identifies the conserved quantity during the time evolution. The left zero mode in covariance space is $\bm l=\widetilde{\bm h}\otimes\widetilde{\bm h}$, thus Lemma 1 gives the conserved quantity
\begin{equation}
\begin{aligned}
\mathcal I
&\equiv\bm l^T\bm{\sigma}(t)\\
&=\widetilde{\bm h}^{\,T}
  \Sigma(t)\widetilde{\bm h}\\
&=\operatorname{Var}\!
  \left[\gamma_1q_1(t)+\gamma_2q_2(t)
        +p_1(t)+p_2(t)\right],
\end{aligned}
\label{eq:coupled_example_invariant}
\end{equation}
where $\operatorname{Var}[\cdot]$ denotes the variance of the variable inside the bracket. This conservation can be checked directly: Eq.~\eqref{eq:coupled_example_eom} gives
\begin{equation}
\frac{d}{dt}
\left(\gamma_1q_1+\gamma_2q_2+p_1+p_2\right)=0.
\label{eq:coupled_example_direct_check}
\end{equation}
In terms of the components of Eq.~\eqref{eq:coupled_example_sigma}, the invariant is
\begin{equation}
\begin{aligned}
\mathcal I={}&
\gamma_1^2\Delta_{q_1}
+\gamma_2^2\Delta_{q_2}
+2\gamma_1\gamma_2\Delta_{q_1q_2}\\
&+2\gamma_1
  \bigl(\Delta_{q_1p_1}+\Delta_{q_1p_2}\bigr)\\
&+2\gamma_2
  \bigl(\Delta_{q_2p_1}+\Delta_{q_2p_2}\bigr)\\
&+\Delta_{p_1}+\Delta_{p_2}
+2\Delta_{p_1p_2}.
\end{aligned}
\label{eq:coupled_example_invariant_expanded}
\end{equation}

Following from Eq.~\eqref{asymptotic_covariance_general}, since there is only one zero mode, the matrices $M_r$ and $M_l$ defined in the proposition in Sec.~\ref{Sec_III} reduce to
\begin{equation}
M_r=\bm r,\qquad M_l=\bm l^T,
\qquad
M_lM_r
=\bigl(\widetilde{\bm h}^{\,T}\bm h\bigr)^2
=(\gamma_1+\gamma_2)^2.
\label{eq:coupled_example_MlMr}
\end{equation}
Equation~\eqref{asymptotic_covariance_general} therefore gives
\begin{equation}
\lim_{t\to\infty}\bm{\sigma}(t)
=
\frac{\mathcal I}
     {(\gamma_1+\gamma_2)^2}\,\bm r.
\label{eq:coupled_example_final_sigma}
\end{equation}
Explicitly, the asymptotic position covariances are
\begin{equation}
\lim_{t\to\infty}\Delta_{q_1}(t)
=
\lim_{t\to\infty}\Delta_{q_2}(t)
=
\lim_{t\to\infty}\Delta_{q_1q_2}(t)
=
\frac{\mathcal I}
     {(\gamma_1+\gamma_2)^2},
\label{eq:coupled_example_final_width}
\end{equation}
while every covariance containing $p_1$ or $p_2$ vanishes.

In this example, two damped particles are connected by a spring. The translationally invariant direction of the landscape and the conserved covariance determine the long-time state in covariance space without requiring an explicit construction of the landscape.  It is interesting to note that, even if one degree of freedom of the two particle system is confined by a quadratic potential, both particles exhibit wave packet localization in the long time limit, which occurs in the single-particle case only when the particle is completely free.

\bibliography{Reference}
\end{document}